# The Importance of Collective Privacy in Digital Sexual and Reproductive Health

Teresa Almeida, Umeå University, Sweden and ITI/LARSyS, Portugal (teresa.almeida@umu.se)
Maryam Mehrnezhad, Royal Holloway University of London, Egham, UK (maryam.mehrnezhad@rhul.ac.uk)
Stephen Cook, Royal Holloway University of London, Egham, UK (stephen.cook.2023@live.rhul.ac.uk)

**Abstract**
There is an abundance of digital sexual and reproductive health technologies that presents a concern regarding their potential sensitive data breaches. We analyzed 15 Internet of Things (IoT) devices with sexual and reproductive tracking services and found this ever-extending collection of data implicates many beyond the individual including partner, child, and family. Results suggest that digital sexual and reproductive health data privacy is both an individual and collective endeavor.

Digital health and the rise of self-tracking technologies have contributed deeply to widening knowledge about sexual and reproductive health, including fertility, family planning, and contraception. These technologies, which include apps and Internet of Things (IoT) devices (such as wearables and linked devices), allow the collection of sensitive data such as sexual activity, menstrual cycle, or pregnancy. While this provides an opportunity to radically impact women's health and mitigate underexplored intimate health issues, it also results in increasing privacy and security risks. Understanding the magnitude of such risks is paramount to addressing concerns that observe, for example, the tracking of abortion[1] and infertility[2]. The gathering of these sensitive data risks bodily autonomy, and can be harmful and potentially life-threatening to the millions involved in (self-) tracking practices when these data are mismanaged, misused, or misappropriated[3]. Yet, it not only has the real-life potential to harm the user of these devices, but these data practices can also impact identified groups including partners (such as partner modes, and sharing functions), baby/child (such as asking about previous pregnancies, births, or about current pregnancies), and friends (through connecting functions, forums, discount links, etc.)[4].

We sought to understand how data privacy is and is not embedded into current digital sexual and reproductive health technologies and how data that is collected on the user implicates not only them but also others, such as their immediate relationships and user groups as a whole - including partners, children, and families. The increasingly popular technologies in our study show how a new set of risks have arisen as various forms of sensors are equipped in these devices and Apps including communicational (WiFi, Bluetooth, NFC), biometric (e.g. fingerprint, face/voice recognition), motion and ambient sensors, the combination of which has not been seen before. Nonetheless, these devices continue to collect data about people's reproductive choices, and sexual activities and gather detailed insights into their intimate lives.

Our results suggest that the design of cyber-safe, privacy-preserving, trustworthy and fair systems and digital products for sexual and reproductive health is paramount. We find out that the Android apps associated with these devices are categorized differently (sometimes multiple categories) on the Android app store including 'intimate', 'health and fitness', 'entertainment', 'habit tracking', 'productivity', 'activity tracking', 'menstruation tracking', 'lifestyle', 'reproductive health', and 'medical'. This miscategorization allows gaming with the data collection since, for example, categorizing an app as 'health and fitness' instead of 'medical' would exempt such an app to comply with the relevant law[3,6].

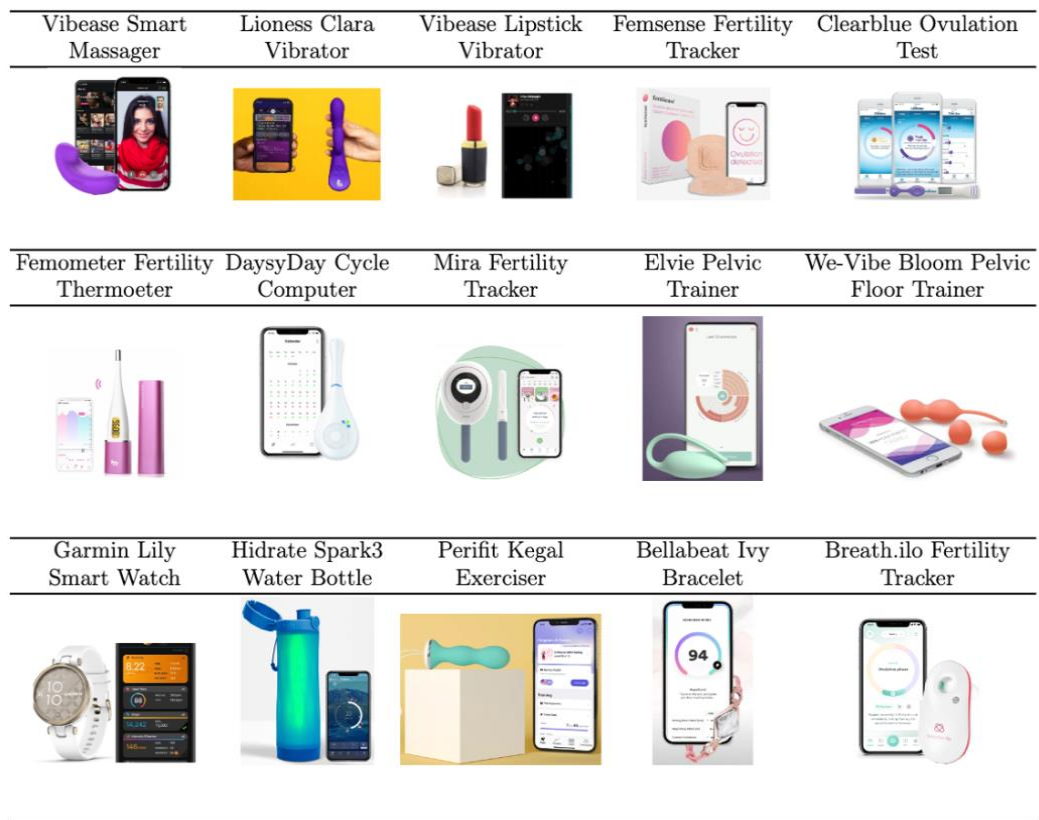

Figure 1: IoT sexual and reproductive health devices that collect data about users and others (partner, child/baby, friends, family). These devices connect to a mobile app and collect data automatically and via user's direct input.

As reported in Table 1, we showed that these devices and apps collect data beyond what needed for their offering including information about users, others, and data about the device and mobile sensors and resources. The identified groups included partners (such as partner modes, sharing functions), baby/child (such as asking about previous pregnancies, births, or about current pregnancies), friends (through connecting functions, forums, discount links, etc.). In these cases, particularly where the collection occurs within the device owner's App, those 'others' about which data is collected have limited agency or ownership over that data. They exist in a liminal space, where they are not the direct 'data subject' under GDPR, but intimate data about them is collected.

These apps have multiple permissions e.g., reading and writing files; where the latter is a highly risky permission allowing to modify system settings. Except for one, all apps request access to location which is seemingly required for BT pairing of the IoT device, though its usage could be irresponsible and violate user privacy. Another highly risky and outdated permission identified in five apps is "GET_ACCOUNT". This permission executes a function that returns a list detailing all accounts on the device, including other users accounts such as email or social media accounts that have previously signed in (e.g., on a password manager app). Similarly, these apps contain several trackers (average: 4.3, min: 1, max: 9) some of which (e.g., Segment and Amplitude) are particularly known for user profiling and tracking.

We also observed that none of these apps present a valid privacy consent when installed and opened for the first time. For instance, the majority of these apps either don't present anything, or bundle the privacy notice with their 'terms and conditions', or include them in their sign-up/sign-in page; all of which are non-compliant practices according to the GDPR[3]. In addition, the privacy policies are beyond the first page and require the user to sometimes go to their website to only face long and difficult to read and understand policies. None of these policies include any content related to data concerning other users than the immediate end user.

Designing digital technologies that contribute to reproductive autonomy, such as those that cater to fertility, family planning, and contraception, without risking individual living and the wider collective is a significant challenge. Overall, data produced within digital sexual and reproductive health enable positive change at the micro-meso-macro level (individual, collective, and societal). But, these data can also leave people more susceptible and vulnerable and it is critical to mitigating risks and the potential for wide-ranging societal implications.

| Product | Vibease Smart Massager | Lioness Clara Vibrator | Vibease Lipstick Vibrator | Femsense Fertility Tracker | Clearblue Ovulation test | We-Vibe Bloom Pelvic Floor Trainer | Breath.ilo Fertility Bracelet | Femometer Fertility Thermoeter | DaysyDay Cycle Computer | Mira Fertility Tracker | Elvie Pelvic Trainer | Garmin Lily Smart Watch | Hidrate Spark 3 Water Bottle | Perifit Kegal Exerciser | Bellabeat Ivy Bracelet |
|---|---|---|---|---|---|---|---|---|---|---|---|---|---|---|---|
| Device | HF | E | HF | HF | HF | LFS | HF | HF | M | HF | HF | HF | HF | HF | HF |
| App Category |  |  |  | HF M RH |  |  |  | HF M RH | MT RH | M |  | HF AC |  |  |  |
| *User's data, sym = symptoms* ||||||||||||||||
| User info | ✓ | ✓ | ✓ | ✓ | ✓ | ✓ | ✓ | ✓ | ✓ | ✓ | ✓ | ✓ | ✓ | ✓ | ✓ |
| Contact info | ✓ | ✓ | ✓ | ✓ | ✓ |  | ✓ | ✓ | ✓ | ✓ | ✓ | ✓ | ✓ | ✓ | ✓ |
| Lifestyle |  |  |  |  |  |  | ✓ | ✓ |  |  |  |  | ✓ | ✓ | ✓ |
| Period |  |  |  | ✓ | ✓ |  | ✓ | ✓ | ✓ | ✓ | ✓ | ✓ |  | ✓ | ✓ |
| Pregnancy |  |  |  | ✓ | ✓ |  | ✓ | ✓ | ✓ | ✓ |  |  | ✓ |  | ✓ |
| Nursing |  |  |  | ✓ |  |  |  |  |  |  |  |  | ✓ |  |  |
| Repro. organs |  | ✓ |  |  | ✓ |  | ✓ | ✓ | ✓ | ✓ |  |  |  | ✓ |  |
| Sex activities | ✓ | ✓ | ✓ | ✓ | ✓ | ✓ | ✓ | ✓ | ✓ | ✓ |  |  |  | ✓ |  |
| Medical |  |  |  | ✓ |  |  |  | ✓ | ✓ | ✓ |  |  |  | ✓ |  |
| Physical sym |  |  |  | ✓ |  |  | ✓ | ✓ | ✓ |  |  | ✓ |  | ✓ | ✓ |
| Emotional sym |  |  |  | ✓ |  |  |  | ✓ | ✓ | ✓ |  |  |  | ✓ | ✓ |
| *Others' data* ||||||||||||||||
| Partner info | ✓ |  | ✓ | ✓ | ✓ | ✓ | ✓ | ✓ | ✓ | ✓ |  | ✓ |  | ✓ | ✓ |
| Social media | ✓ | ✓ | ✓ |  |  |  | ✓ | ✓ | ✓ | ✓ |  | ✓ | ✓ | ✓ | ✓ |
| Child info |  |  |  | ✓ | ✓ |  | ✓ | ✓ |  |  | ✓ | ✓ |  |  |  |
| *IoT/Mobile device's resources* ||||||||||||||||
| Storage | ✓ | ✓ | ✓ |  |  | ✓ | ✓ | ✓ | ✓ | ✓ |  | ✓ | ✓ | ✓ | ✓ |
| Call/Contacts | ✓ |  | ✓ |  |  |  |  |  | ✓ |  |  | ✓ | ✓ | ✓ |  |
| Calendar |  |  |  |  |  |  |  |  |  |  |  |  | ✓ |  |  |
| WiFi | ✓ | ✓ | ✓ | ✓ | ✓ | ✓ | ✓ | ✓ | ✓ | ✓ | ✓ | ✓ | ✓ | ✓ | ✓ |
| Cam/Mic | ✓ | ✓ | ✓ |  |  |  |  | ✓ | ✓ | ✓ |  | ✓ |  |  |  |
| GPS | ✓ | ✓ | ✓ |  | ✓ | ✓ | ✓ | ✓ | ✓ | ✓ | ✓ | ✓ |  | ✓ | ✓ |
| Bluetooth | ✓ | ✓ | ✓ |  | ✓ | ✓ | ✓ | ✓ | ✓ | ✓ | ✓ | ✓ | ✓ | ✓ | ✓ |
| Sensor data | ✓ | ✓ |  | ✓ |  |  | ✓ | ✓ | ✓ | ✓ | ✓ | ✓ | ✓ | ✓ | ✓ |
| NFC |  |  |  | ✓ |  |  |  |  |  |  |  |  | ✓ |  |  |

Table 1: User data collected by FemTech IoT Devices. Google Play Store App Categories are shown as Health and Fitness = HF, Entertainment = E, Lifestyle = LS, Habit Tracking = HT, Productivity = Prod, Activity Tracking = AC, Menstruation Tracking = MT, Reproductive Health = RH, Medical = M.

**Method**

**IoT device/app set and data collection:** We utilize a security and privacy method to conduct an analysis of the data collection and sharing of these devices. We chose a range of FemTech devices advertised for digital sexual and reproductive health (Figure 1), including devices that monitor fertility, those that track and predict menstrual cycles, and those that collect such data directly from the user as a part of their wider offerings. We set up these devices (turning on and connecting to their associated Android App), and used them like an end user. These devices might have different ways of measuring the fertile window (e.g., sleep, temperature, $CO_2$). There are also devices which are not directly advertised for sexual and reproductive health (e.g., Hydrate Smart Bottle), but collect data about period, pregnancy, and nursing. We particularly focus on devices which not only collect extensive data about the user, but also others such as child/baby, partner, and family/friend via either asking directly and/or having access to one's social media or phone calls and contacts.

**Security and privacy analysis**: We utilize Exodus[7] (a privacy auditing platform for android apps) to identify the embedded trackers and permissions that are present in each app. Exodus analyzes the app code via a static analysis method to report the permissions and trackers. Additionally, while installing these apps, we observed whether or not they present the user with any privacy notice and/or consent. We also went through the privacy policies (on the app and via links on their websites) and particularly looked for information regarding collective privacy. This was a manual process and was repeated by two of the authors to assure consistency in the results.

## Data Availability
The list of the devices used in our experiments is presented in the table and hence publicly available. We set up a dummy account and the data given to each device and app was random. Our methods can be duplicated by others on this set of devices. Full tables reporting the practices of each device regarding app permission lists and built-in trackers are available under request.

## Acknowledgments
This work has been supported by the EU Horizon 2020 research and innovation program under grant agreement number 952226 and the PETRAS National Centre of Excellence for IoT Systems Cybersecurity, which has been funded by the UK EPSRC under grant number EP/S035362/1.This work has been supported by the EU Horizon 2020 research and innovation program under grant agreement number 952226 and the PETRAS National Centre of Excellence for IoT Systems Cybersecurity, which has been funded by the UK EPSRC under grant number EP/S035362/1.

## Author Contributions
All authors contributed to the writing and review of the manuscript. Stephen Cook and Maryam Mehrnezhad performed the security analysis of the apps and IoT devices in Feb 2023. All authors approved the final version of the manuscript.

## References

[1] Friedman, A. B., Bauer, L., Gonzales, R., & McCoy, M. S. (2022). Prevalence of Third-Party Tracking on Abortion Clinic Web Pages. JAMA Internal Medicine, 182(11), 1221-1222. 10.1001/jamainternmed.2022.4208
[2] Costa Figueiredo, M. (2021). Data Work and Data Tracking Technologies in Fertility Care: A Holistic Approach (Order No. 28542620). Available from ProQuest Dissertations & Theses Global. (2558053193). Retrieved from http://proxy.ub.umu.se/login?url=https://www.proquest.com/dissertations-theses/data-work-tracking-technologies-fertility-care/docview/2558053193/se-2
[3] Mehrnezhad, M. and Almeida, T. Caring for Intimate Data in Fertility Technologies. (2021) CHI Conference on Human Factors in Computing Systems (Yokohama, Japan) (CHI '21). Association for Computing Machinery, New York, NY, USA, Article 409, 11 pages. https://doi.org/10.1145/ 3411764.3445132
[4] Mehrnezhad, M., Shipp, L., Almeida, T., and Toreini, E. (2022). Vision: Too Little Too Late? Do the risks of FemTech already outweigh the benefits?. In 2022 European Symposium on Usable Security (EuroUSEC 2022)
[5] Almeida, T., Shipp, L., Mehrnezhad, M., and Toreini, E. (2022). Bodies Like Yours: Enquiring Data Privacy in FemTech. Adjunct Proceedings of the 2022 Nordic Human-Computer Interaction Conference. 2022.
[6] McMillan, C. (2023). Rethinking the Regulation of Digital Contraception Under the Medical Devices Regime." Medical Law International: 09685332231154581.
[7] https://exodus-privacy.eu.org/en/